\documentclass[lettersize,journal]{IEEEtran}
\usepackage{amsmath,amsfonts}
\usepackage{array}
\usepackage[caption=false,font=normalsize,labelfont=sf,textfont=sf]{subfig}
\usepackage{textcomp}
\usepackage{stfloats}
\usepackage{url}
\usepackage{verbatim}
\usepackage{graphicx}
\def\BibTeX{{\rm B\kern-.05em{\sc i\kern-.025em b}\kern-.08em
    T\kern-.1667em\lower.7ex\hbox{E}\kern-.125emX}}
\usepackage{balance}

\usepackage[mode=buildmissing]{standalone}
\usepackage{acro}
\DeclareAcronym{AE}{
short=AE,
long= autoencoder,
}

\DeclareAcronym{WDM}{
short=WDM,
long=wavelength division multiplexing ,
}

\DeclareAcronym{NN}{
short=NN,
long= neural network,
}

\DeclareAcronym{PS}{
short= PS,
long= pulse-shaping,
}

\DeclareAcronym{DAC}{
short=DAC,
long= digital-to-analog converter,
}

\DeclareAcronym{ADC}{
short=ADC,
long= analog-to-digital converter,
}

\DeclareAcronym{DL}{
short=DL,
long= deep learning,
}

\DeclareAcronym{ENOB}{
short=ENOB,
long= effective number of bit,
}

\DeclareAcronym{DPD}{
short= DPD,
long= digital pre-distortion,
}

\DeclareAcronym{}{
short=,
long= ,
}

\DeclareAcronym{MZM}{
short= MZM,
long= Mach-Zehnder modulator ,
}

\DeclareAcronym{MZI}{
short= MZM,
long= Mach-Zehnder interferometer ,
}

\DeclareAcronym{EDFA}{
short= EDFA,
long= erbium-doped fiber amplifier ,
}

\DeclareAcronym{SE}{
short= SE,
long= spectral efficiency ,
}

\DeclareAcronym{PA}{
short= PA,
long= power amplifier ,
}

\DeclareAcronym{MF}{
short= MF,
long= matched filter ,
} 

\DeclareAcronym{OOB}{
short= OOB,
long= out-of-band ,
}

\DeclareAcronym{ICI}{
short= ICI,
long= inter-channel interference,
}

\DeclareAcronym{IQM}{
short= IQM,
long= in-phase and quadrature modulator,
}

\DeclareAcronym{RRC}{
short= RRC,
long= root-raised-cosine,
}

\DeclareAcronym{AWGN}{
short= AWGN,
long= additive white Gaussian noise,
}

\DeclareAcronym{SER}{
short= SER,
long= symbol error rate,
}

\DeclareAcronym{RL}{
short= RL,
long= reinforcement learning,
}

\DeclareAcronym{SNDR}{
short= SNDR,
long= signal-to-noise-plus-distortion ratio,
}

\DeclareAcronym{SNR}{
short= SNR,
long= signal-to-noise ratio,
}

\DeclareAcronym{FIR}{
short= FIR,
long= finite impulse response,
}

\usepackage{amsmath} 
\usepackage{amssymb}
\usepackage{amsthm}
\usepackage{bm}
\usepackage{color}
\usepackage{verbatim}
\usepackage{booktabs}
\usepackage{amsmath}
\usepackage{balance}
\usepackage{graphicx}
\usepackage{tabu}
\usepackage[font={footnotesize}]{caption}
\usepackage{array, ltablex, multirow,ragged2e}
\usepackage{hyperref}
\usepackage{longtable}
\usepackage{mdframed}
\usepackage{subfig} 
\usepackage{dsfont}
\usepackage{cite}
\usepackage{float}
\usepackage{algorithm}
\usepackage{amsmath}
\usepackage{amssymb}

\usepackage[noend]{algpseudocode}
\usepackage{tikz,pgfplots}
\usepgfplotslibrary{fillbetween}
\usepgfplotslibrary{groupplots}
\pgfplotsset{compat=newest}%
\usetikzlibrary{positioning,matrix,shapes.multipart,shapes.misc,spy}%
\tikzstyle{annotation}=[fill=white]%

\newcommand{\vect}[1]{\boldsymbol{#1}}

\newcommand{\Ctranspose}{\mathsf{H}}

\DeclareMathOperator*{\argmax}{arg\,max}

\usetikzlibrary{shapes.arrows}
\usetikzlibrary{spy}

\tikzset{%
	partial ellipse/.style args={#1:#2:#3}{%
		insert path={+ (#1:#3) arc (#1:#2:#3)}%
	}%
}%

\tikzstyle{style A}=[black, mark=diamond*, mark options={solid, fill=white, mark size=2.0pt}, solid]%
\tikzstyle{style B}=[color=blue, mark=*, mark options={solid, fill=white, mark size=1.5pt}, dotted]%
\tikzstyle{annotation}=[fill=white]%

\tikzstyle{star marker}=[mark=star, mark options={solid, scale=1}]
\tikzstyle{diamond marker}=[mark=diamond*, mark options={solid, fill=white, mark size=2.0pt}]
\tikzstyle{triangle marker}=[mark=triangle*, mark options={solid, scale=0.8}]
\tikzstyle{square marker}=[mark=square*, mark options={solid, scale=0.8}]
\tikzstyle{circle marker}=[mark=*, mark options={solid, scale=1}]

\tikzstyle{autoencoder}=[thick, color=red, triangle marker, dashed]%
\tikzstyle{baseline}=[thick, color=blue, square marker, solid]%
\tikzstyle{ibaseline}=[thick, color=green!50!black, no markers, solid]%

\makeatletter
\def\BState{\State\hskip-\ALG@thistlm}
\makeatother

\theoremstyle{plain}

\theoremstyle{remark}

\theoremstyle{plain}

\graphicspath{{../figures/}}

\newcommand{\JS}[1]{{\color{black}{#1}}}

\begin{document}

\title{\title{Model-Based End-to-End Learning for WDM \\ Systems  With Transceiver Hardware Impairments}
}

\author{Jinxiang Song, \emph{Student Member, IEEE},
Christian H\"{a}ger, \emph{Member, IEEE}, 
Jochen Schr\"{o}der, \emph{Member, IEEE}, \\
Alexandre Graell i Amat, \emph{Senior Member, IEEE}, 
and Henk Wymeersch, \emph{Senior Member, IEEE}
\thanks{Parts of this paper have been presented at the \emph{ Optical Fiber Communication Conference and Exhibition (OFC)}, San Diego, California, USA, 2021 \cite{song2021end}.}%
\thanks{This work was supported by the Knut and Alice Wallenberg Foundation, grant No.~2018.0090, and the Swedish Research Council under grant  No.~2018-0370. 
\emph{(Corresponding author: Jinxiang Song)}}
\thanks{%
Jinxiang Song, Christian H\"{a}ger, Alexandre Graell i Amat, and Henk Wymeersch are with the Department of Electrical Engineering, Chalmers University of Technology, 41296 Gothenburg, Sweden (emails: \{jinxiang, christian.haeger, alexandre.graell, henkw\}@chalmers.se).}
\thanks{%
Jochen Schr\"{o}der is with the Department of Microtechnology and Nanoscience, Chalmers University of Technology, 41296 Gothenburg, Sweden (email: jochen.schroeder@chalmers.se)}
}


\IEEEspecialpapernotice{(Invited Paper)}

\maketitle

\begin{abstract}
We propose an \ac{AE}-based transceiver for a \ac{WDM} system impaired by hardware imperfections. We design our \ac{AE} following the architecture of conventional communication systems. This enables  to initialize the AE-based transceiver to have similar performance to its conventional counterpart prior to training and  improves the training convergence rate. We first train the \ac{AE} in a single-channel system, and show that it achieves performance improvements by putting energy outside the desired bandwidth, and therefore cannot be used for a \ac{WDM} system.
We then train the \ac{AE} in a \ac{WDM} setup. Simulation results show that the proposed \ac{AE} significantly outperforms the conventional approach. More specifically, it increases the spectral efficiency of the considered system by reducing the guard band by $37\%$ and $50\%$ for a root-raised-cosine filter-based matched filter with $10\%$ and $1\%$ roll-off, respectively. An ablation study indicates that the performance gain can be ascribed to the optimization of the symbol mapper, the pulse-shaping filter, and the symbol demapper. Finally, we use reinforcement learning to learn the pulse-shaping filter under the assumption that the channel model is unknown. Simulation results show that the reinforcement-learning-based algorithm achieves similar performance to the standard supervised end-to-end learning approach assuming perfect channel knowledge.
\end{abstract}


\section{Introduction}
The ever-growing demand for higher data rates drives the rapid development of optical fiber communication systems. One of the most important developments is \acf{WDM} transmission, where parallel data channels are transmitted on different wavelengths simultaneously. The throughput of modern \ac{WDM} systems often exceeds tens of $\mathrm{Tb/s}$ with more than $100$ channels~\cite{winzer2018fiber}. However, the overall bandwidth of fiber systems is limited by the bandwidth of \acp{EDFA} that periodically amplify the signals along the fiber link~\cite{yamada1998gain}. Optimizing the \acf{SE}, i.e., the number of bits that can be transmitted per unit time and frequency, is therefore crucial to further increase the throughput of fiber optical systems.

Over the last decade, most works have focused on increasing the per-channel \ac{SE} via advanced modulation formats using coherent detection. The fiber nonlinearity and hardware impairments, such as the \acp{ENOB} of the \acf{DAC}, however, severely limit the per-channel \ac{SE}. Furthermore, spectrum gaps between individual channels, which are often referred to as guard bands, waste significant bandwidth and limit the overall system throughput. Hence, the guard bands between channels need to be minimized. 
The most promising solution has been the application of flexible grids, which allows for transmission with flexible channel bandwidths thus enabling simultaneous transmission of mixed bit rates~\cite{rafique2013flex} and allowing to reduce \ac{SE} loss from guard bands for optical filtering. 

\JS{To minimize the guard bands between channels, it is common to employ pulse shaping to create a near-rectangular spectrum in the frequency domain with a bandwidth close to the symbol rate. However, in practice, generating a rectangular spectrum is difficult due to the finite  \ac{PS} filter and transceiver hardware impairments, requiring computation expensive digital signal processing to eliminate performance degradation caused by \ac{ICI} \cite{mazur2019joint}. Guard bands therefore remain a major contributor to \ac{SE} loss in \ac{WDM} systems.}

\begin{table*}[t!]
\setlength{\tabcolsep}{0.6em}
\scriptsize
\centering
\vspace{0.15cm}
\caption{Applications of end-to-end AE-learning in communication systems }
\begin{tabular}{c|c|c|c|c|c|c|c|c}
\toprule
&Ref. & year &Application   & isolated ch. & ICI ch. & sim. &exp. & Description\\
\midrule
\multirow{15}{*}{Wireless} & \cite{o2017introduction}    & 2017  & geom. shaping & $\checkmark$ &           & $\checkmark$ &        & \ const. mapper/demapper training over an AWGN channel \\ 

 & \cite{kim2017novel}    & 2017  & geom. shaping & $\checkmark$ &           &$\checkmark$    &        & const. mapper/demapper training for PAPR reduction \\ 
 
 & \cite{o2017physical}    & 2017  & geom. shaping \& precoding &  &    $\checkmark$       & $\checkmark$   &        &  MIMO precoding/decoding \\

                           & \cite{dorner2017deep}      & 2017  & geom. shaping & $\checkmark$ &           & $\checkmark$          & $\checkmark$  &mapper/demapper training in sim., demapper tuning in exp. \\ 
                           
                           & \cite{felix2018ofdm}      & 2018  & geom. shaping & $\checkmark$ &                   & $\checkmark$   &  & mapper/demapper training for OFDM system \\ 
                           
                           & \cite{ye2018channel}      & 2018  & geom. shaping & $\checkmark$ &                   & $\checkmark$   &  & mapper/demapper training over a learned channel via GAN \\ 
                           
                           & \cite{aoudia2018end}      & 2018  & geom. shaping & $\checkmark$ &           & $\checkmark$          & $\checkmark$  &mapper/demapper training without knowing the channel model \\ 
                            & \cite{zhang2019neural}      & 2019  & geom. shaping & $\checkmark$ &           & $\checkmark$          &   & mapper/demapper training for PAPR reduction \\ 
                           
                            & \cite{choi2019neural}      & 2019  & joint channel/source coding & $\checkmark$ &                   & $\checkmark$   &  & Joint channel and source coding/decoding \\ 
                            
                           & \cite{stark2019joint}    & 2019  & geom. \& prob. shaping   & $\checkmark$ &  & $\checkmark$ & & Joint geom. and prob. shaping/demapping \\ 
                            & \cite{cammerer2020trainable}    & 2020  & geom. shaping \& coding   & $\checkmark$ &  & $\checkmark$ & $\checkmark$  & mapper/demapper learning and error correction code design \\ 
                            
                            & \cite{van2020deep}    & 2020  & geom. shaping    & $\checkmark$ &  & $\checkmark$ & & mapper/demapper training for OFDM and multi-user system \\ 
                        
                             & \cite{aoudia2021end}    & 2021  & geom. shaping    & $\checkmark$ &  & $\checkmark$ & & mapper/demapper training for OFDM system \\ 
           
                            & \cite{aoudia2021waveform}    & 2021  & geom. shaping \& waveform   & $\checkmark$ & $\checkmark$  & $\checkmark$ & & Joint transceiver training \\

\midrule                            
\multirow{10}{*}{Fiber optic} & \cite{karanov2018end}    & 2018  &  geom. shaping \& waveform & $\checkmark$ & & $\checkmark$ & $\checkmark$  &Joint tranceiver learning for IM/DD system   \\ 
                           & \cite{li2018achievable}     & 2018  & geom. shaping &  $\checkmark$ &  & $\checkmark$ & & mapper/demapper training for the nonlinear fiber channel \\ 
                            & \cite{jones2018geometric}    & 2018  & geom. shaping & $\checkmark$ &  & $\checkmark$ & & mapper/demapper training for the fiber channel \\ 
                             & \cite{jones2019end}    & 2019  & geom. shaping & $\checkmark$ &  & $\checkmark$ & & mapper/demapper training for optimizing GMI \\ 
                             
                             & \cite{karanov2019end}    & 2019  &  geom. shaping \& waveform & $\checkmark$ &  & $\checkmark$ & $\checkmark$ & Joint transceiver learning for IM/DD system \\ 
                             
                             & \cite{gumucs2020end}    & 2020  & geom. shaping & $\checkmark$ &  & $\checkmark$ & & mapper/demapper training for optimizing GMI\\
                              & \cite{uhlemann2020deep}    & 2020  & geom. shaping \& waveform  & $\checkmark$ &  & $\checkmark$ & & Joint transceiver learning for single channel transmission\\
                              
                             & \cite{karanov2020end}    & 2020  & Prob. shaping  & $\checkmark$ &  & $\checkmark$ & &  Prob. shaping  \\ 
                            & \cite{jovanovic2021end}    & 2021  & geom. shaping & $\checkmark$ &  & $\checkmark$ & & mapper/demapper training for varying SNR and laser linewidth\\ 
                            & \cite{song2021end}    & 2021  & geom. shaping \& waveform  & $\checkmark$ & $\checkmark$ &   $\checkmark$ & & Joint transceiver design for superchannel systems \\ 
                             & This work    & 2021  & geom. shaping \& waveform  & $\checkmark$ & $\checkmark$ &   $\checkmark$ & & Joint transceiver design for densely-spaced \ac{WDM} systems \\ 
\bottomrule
\end{tabular}
\RaggedRight

\smallskip
isolated ch.: the channel does not suffer from \ac{ICI}; \ac{ICI} ch.: channel that suffers from \ac{ICI}.
\label{tab:applications of AE}
\end{table*}

In recent years, the rapid improvement of machine learning techniques has led to a resurgence of interest in applying \acl{DL} techniques for communication systems~\cite{jiang2016machine, khan2017machine}. Most work has focused on supervised learning for a \emph{specific functional block}, e.g.,  modulation recognition~\cite{west2017deep}, carrier recovery~\cite{zibar2015machine}, and fiber nonlinearity mitigation~\cite{haeger2018}, with the aim of finding better performing (or less complex) algorithms by replacing the conventional model-based methods with \acp{NN}. In contrast to focusing on specific functional blocks, \emph{end-to-end learning} has been proposed to design the transmitter and receiver jointly\cite{o2017introduction}. The key idea is to interpret the transceiver design as a reconstruction task, whereby the transmitter and the receiver can be implemented as an \acf{AE} and thus jointly optimized. This method has led to several applications for both wireless~\cite{o2017introduction,dorner2017deep,felix2018ofdm} and optical communications~\cite{li2018achievable, jones2018geometric, jovanovic2021end}.  A broad, but non-exhaustive overview of existing work is listed in Table~\ref{tab:applications of AE}. We observe that (i) a majority of works relate to wireless rather than optical communication; (ii) geometric constellation shaping for different channels and applications has been the main focus; (iii) there are few experimental validations. Studies that also learn waveforms and equalizers are limited to~
\cite{karanov2018end,uhlemann2020deep,aoudia2021waveform}. In \cite{karanov2018end}, the whole transceiver is implemented as an \ac{AE}, and transmission is demonstrated over a short-haul intensity modulation/direct detection (IM/DD) system. However, the \ac{NN} in~\cite{karanov2018end} is used as a ``black-box'' and it is difficult to interpret the learned solution.
In \cite{uhlemann2020deep}, the transmitter is implemented as a trainable constellation mapper combined with a trainable \ac{PS} filter, and it is shown that the PS filter can be learned to mitigate chromatic dispersion and Kerr effect. An explicit low-pass filter is used to reduce information loss and thus avoid \ac{OOB} emissions. A related approach has recently been applied in~\cite{aoudia2021waveform}, where flexible constellations and waveforms for wireless dispersive channels under \ac{OOB} power leakage constraints were learned.  

In this paper, we apply end-to-end learning to an multi-channel \ac{WDM} system. Similar to\cite{uhlemann2020deep,aoudia2021waveform} (for a single-channel system), we consider designing several transceiver blocks \textemdash constellation mapper, \ac{PS} filter, \acf{DPD}, and demapper\textemdash jointly. Such an \ac{AE} design incorporates the expert domain knowledge of conventional communication systems and therefore allows for (i) training speed improvements via meaningful parameter initialization and (ii) performance gain explanation through an ablation study.  The main contributions of this paper are:
\begin{itemize}
    \item We propose a novel end-to-end \ac{AE} for \ac{WDM} transceivers with non-ideal \ac{DAC} and \ac{IQM}. We decompose the transmitter \ac{NN} into a concatenation of small (simple) \acp{NN},  each corresponding to a functional block of a conventional communication system. Our approach differs from \cite{uhlemann2020deep,aoudia2021waveform}  in terms of the considered hardware impairments and how \ac{OOB} emissions are accounted for: instead of a low-pass filter~\cite{uhlemann2020deep} or a constraint\cite{aoudia2021waveform}, we  show that the \ac{AE} automatically learns to avoid/adapt \ac{OOB} emissions to minimize the end-to-end loss. 
   
    \item \JS{We highlight the potential pitfalls when using end-to-end \ac{AE}-learning for designing hardware-impaired communication systems. In particular, we show that when the \ac{AE} is trained for a single-channel system, it achieves performance improvements by putting energy outside the desired signal bandwidth, which would cause large \ac{ICI} in \ac{WDM} systems when the channels are closely spaced. We demonstrate that if the \ac{AE} is instead trained with three channels, it learns to limit \ac{ICI} while still outperforming the considered baseline. However, care must be taken for the sampling rates or bandwidths used during training to match experimental constraints to avoid unrealistic gains.}
   
     \item We conduct a thorough ablation study and show that the performance improvement of the AE-based system is ascribed to the optimization of the constellation mapper, the \ac{PS} filter, and the demapper. Therefore, we show that our proposed method increases the interpretability compared to conventional AE-based systems. Additionally, we provide reproducible open-source implementations of our \acp{AE} and benchmark scheme.\footnote{The complete source code to reproduce all results in this paper is available at \url{https://github.com/JSChalmers/AE-Based-WDM-Transceivers}.}
    
    \item We extend the model-free training algorithm proposed in~\cite{aoudia2018end, aoudia2019model}, so that the \ac{RL} based transmitter training algorithm can be applied to train the \ac{PS} filter, for which memory effects need to be considered. The resulting training algorithm is shown to achieve similar performance to the standard end-to-end learning approach assuming a perfect channel model. 
    This opens the door toward experimental implementation of the proposed \ac{AE}.
   
\end{itemize}

The remainder of this paper is structured as follows. In Section~\ref{background}, we give a brief introduction to DL basics and the concept of AE-based communications. Then, in Section~\ref{superchannel}, we introduce the generic setup of closely-spaced \ac{WDM} systems and the main hardware limitations. Section~\ref{proposed_AE} introduces the proposed AE-based \ac{WDM} system and simulation results are provided in Section~\ref{results}. Finally, the paper is concluded in Section~\ref{conclusion}.

\subsubsection*{Notation}
$\mathbb{Z}$, $\mathbb{R}$, and $\mathbb{C}$ denote the sets of integers, real numbers, and complex numbers, respectively. Column vectors will be denoted with lower case letters in bold (e.g., $\vect{x}$), with $x_n$ 
referring to the $n$-th entry in $\vect{x}$, and $\vect{x}_{n}^{(L)}$ denotes the column vector consisting of $(n-L)$-th to $(n+L)$-th elements of $\vect{x}$; $\lvert\cdot\rvert$ returns the absolute value of a real number, and $\lvert \Im\{\vect{x} \}\rvert$ and $\lvert \Im\{\vect{x} \}\rvert$  return the absolute value of the real and imaginary part of each element in $\vect{x}$, respectively; $(\cdot)^\top$ and $(\cdot)^\Ctranspose$ denote transpose and conjugate transpose, respectively.  Matrices will be denoted in bold capitals (e.g., $\vect{X}$), and $\vect{I}_{N}$ denotes identity matrix of size $N$; $[a,b]^M$ is the $M$-fold Cartesian product of the interval $[a,b]$. Lastly, $\mathbb{E}\{\cdot\}$ denotes the expectation operator.

\section{Deep Learning and Autoencoder-Based Communication systems}
\label{background}

In this section, we start by reviewing the general theory behind \ac{DL}, followed by a brief introduction to the concept of AE-based communication systems. Then, we introduce the training of AE-based communication systems under two assumptions: (i) the channel model is known and differentiable, and (ii) the channel model is unknown or not differentiable.

\subsection{Neural Networks and Gradient-Based Learning}

\begin{figure*}[t]
    \centering
    \includegraphics[width=1\textwidth]{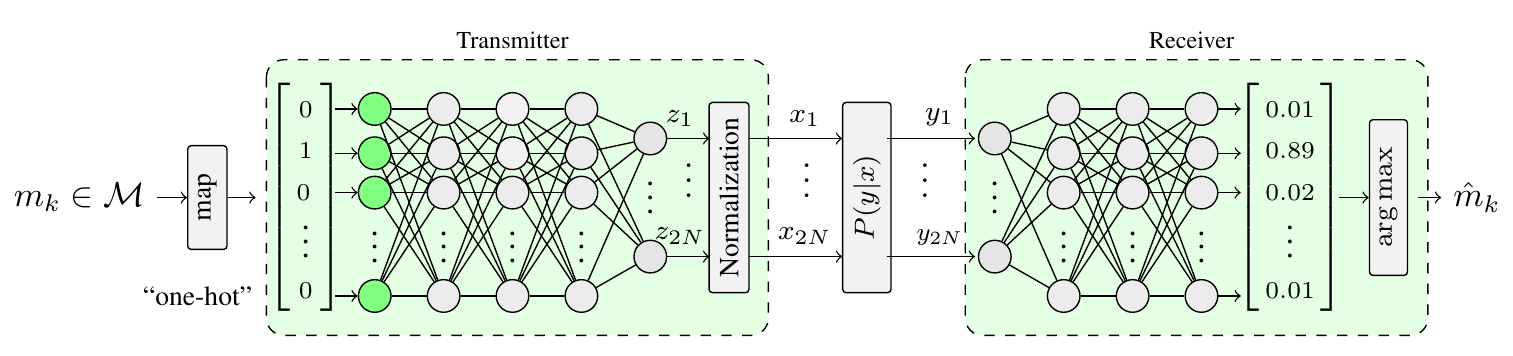}
    \caption{ Example of an AE-based communication system, where the transmitter and receiver are implemented by a pair of fully connected NNs.}
    \label{fig:AE}
\end{figure*}

\subsubsection{Feedforward NN}
A feedforward \ac{NN} with $K$ layers is a parametric function $f(\vect{r}_0; \vect{\theta}): \mathbb{R}^{N_0} \to \mathbb{R}^{N_K}$ that maps  an input vector $\vect{r}_0 \in \mathbb{R}^{N_0}$ to an output vector $\vect{r}_K \in \mathbb{R}^{N_K}$ through $K$ sequential processing steps according to
\begin{align}
    \vect{r}_k = f_k (\vect{r}_{k-1}; \vect{\theta}_k), \quad k=\{1,\ldots, K\}\,, 
\end{align}
where  $f_k(\vect{r}_{k-1}; \vect{\theta}_k): \mathbb{R}^{N_{k-1}} \to \mathbb{R}^{N_k}$ is the mapping carried out by the $k$-th layer. Here, the mapping of the $k$--th layer is defined by the set of parameters $\vect{\theta}_k$, and the entire \ac{NN} is defined by $\vect{\theta} = \{\vect{\theta}_1,\ldots, \vect{\theta}_K \}$. A commonly used type of feedforward \ac{NN} is the fully connected \ac{NN} in which all layers have the form 
\begin{align}
    f_k (\vect{r}_{k-1}; \vect{\theta}_k) = \sigma(\vect{W}_k \vect{r}_{k-1} + \vect{b}_k)\,,
\end{align}
where $\vect{W}_k\in \mathbb{R}^{N_{k-1} \times N_k }$ is a weight matrix, $\vect{b}_k \in \mathbb{R}^{N_k}$ is a bias vector, and $\sigma (\cdot)$ is a point-wise \emph{activation} function. Hence, the set of trainable parameters of the $k$-th layer is $\vect{\theta}_k = \{\vect{W}_k, \vect{b}_k\}$. An example of a fully connected  \ac{NN} is shown in the transmitter and the receiver in Fig.~\ref{fig:AE}.

\subsubsection{Gradient-based learning}

Training of the \ac{NN} can be performed in an iterative fashion with data-driven gradient-based optimization methods. Given a set of labeled training data $\mathcal{D}\subset \{\mathcal{X}\times \mathcal{Y}\}$, where $\mathcal{X}$ and $\mathcal{Y}$ are the input and output alphabets, the training objective is to find the set of parameters $\vect{\theta}$ such that the average loss
\begin{align}
    \mathcal{L}_{\mathcal{D}}(\vect{\theta}) =\frac{1}{|\mathcal{D}|} \sum_{(x,y)\in \mathcal{D}} \ell(f(x; \vect{\theta}),y)
\end{align}
between the \ac{NN} output $\hat{y} = f(x, \vect{\theta})$ and the true label $y\in\mathcal{Y}$ is minimized. Here, $|\mathcal{D}|$ is the size of the training data set and $\ell (f(x;\vect{\theta}),y)$ is the per-example loss function associated with
returning the output $ \hat{y}= f(x; \vect{\theta})$ when $y$ is the true label. In practice, when the training data set $\mathcal{D}$ is large, computing the gradients of the average loss over the whole training data set is computationally expensive, and the parameter set $\vect{\theta}$ is commonly optimized by using stochastic gradient descent (SGD) or its variants as follows. For each training iteration $t$, a minibatch $\mathcal{B}_t$ is sampled from $\mathcal{D}$. Then, the parameter set $\vect{\theta}$ is updated according to 
\begin{align}
    \label{eq:SGD}
    \vect{\theta}_{t+1}= \vect{\theta}_t - \alpha \nabla_{\vect{\theta}}\mathcal{L}_{\mathcal{B}_t}(\vect{\theta}_t),
\end{align}
where $\alpha > 0$ is the learning rate.
In practice, SGD sometimes suffers from slow convergence rate due to problems like small gradients at suboptimal values of $\vect{\theta}$. To improve the convergence rate of SGD, many variants of SGD using momentum~\cite{wilson2017marginal} or adaptive learning rate~\cite{zeiler2012adadelta} have been proposed.

\subsection{End-to-End \ac{AE} Learning-Based Communication Systems}

\subsubsection{AE-based Communication Systems} 
\label{AE_mapping}
End-to-end learning of AE-based communication systems  was originally proposed  in~\cite{o2017introduction}, where the transceiver for a given channel with  channel law $p(\vect{y}|\vect{x})$  is implemented by  a  pair  of  \acp{NN} $f_{\vect{\tau}}: \mathcal{M}\to \mathbb{C}^{N}$ and $f_{\vect{\rho}}: \mathbb{C}^{N} \to [0,1]^M $. Here, $\mathcal{M}=\{1,\ldots,M\}$ is the message set, $N$ is the number of complex channel uses, and $\vect{\tau}$ and $\vect{\rho}$ are the sets of trainable \ac{NN} parameters. Fig.~\ref{fig:AE} depicts the general setup of an AE-based communication system.

\emph{Transmitter:}
Given a message $m_k \in \mathcal{M}$, it is first encoded as an $M$-dimensional ``one-hot'' vector, where the $m_k$-th element is $1$ and all the others are $0$.\footnote{The ``one-hot'' encoding is the standard way of representing categorical values in most machine learning algorithms\cite{goodfellow2016deep} and
facilitates the minimization of the \ac{SER}. However, the dimension of the ``one-hot'' vector grows exponentially with the number of bits in each message and therefore increases the \ac{NN} size. Alternative embeddings~\cite{rodriguez2018beyond} and multi-hot sparse categorical cross-entropy loss can be used to alleviate this problem.} Then, the transmitter \ac{NN} takes this ``one-hot'' vector as input and generates a vector of $2 N$ outputs $\vect{x}_k = f_{\vect{\tau}}(m_k)$, where the $2 N$ outputs correspond to the real and imaginary part of the transmitted vectors. The average transmit power constraint $\mathbb{E}\{||\vect{x}_k||^2\}\leq N P_T$, where $P_T$ is the average transmit power per channel use, is enforced by a normalization layer~\cite{o2017introduction}.

\emph{Receiver:}
\label{AE_Rx}
The symbol $\vect{x}_k$ is sent over the channel in $N$ complex channel uses, after which $\vect{y}_k$ is observed at the receiver.  The receiver \ac{NN} processes the received vector $\vect{y}_k$ by generating an $M$-dimensional probability vector $\vect{q}_k=f_{\vect\rho}(\vect{y}_k)$, where the components of $\vect{q}_k$ can be interpreted as the estimated posterior probabilities of the messages. 
Finally, the transmitter generates the estimate of the transmitted message according to $\hat{m}_k=\argmax_m[\vect{q}_k]_m$, where $[\vect{x}]_m$ returns the $m$-th element of $\vect{x}$.

\subsubsection{End-to-End Training With a Known Channel Model}

To optimize the transmitter and receiver parameters, it is crucial to have a suitable optimization criterion. Due to the fact that the optimization relies on the empirical computation of gradients, a criterion like block error rate (BLER), i.e., $\mathrm{Pr}\{\hat{m}_k\neq m_k\}$, cannot be used directly (as the BLER is not differentiable).
Instead, a commonly used criterion is the cross-entropy loss~\cite{o2017introduction}, defined by
\begin{equation}
\label{eq:ce_loss}
    \mathcal{L}(\vect{\tau}, \vect{\rho})=-\mathbb{E}\{ \log [f_{\vect{\rho}}(\vect{y}_k)]_{m_k}\},
\end{equation}
where the dependence of $\mathcal{L} (\vect{\tau},\vect{\rho})$ on $\vect{\tau}$ is implicit through the distribution of the channel output $\vect{y}_k$, which is a function of the channel input $f_{\vect{\tau}}(m_k)$.

The transmitter and receiver parameters are optimized in an iterative fashion as follows. In each training iteration $t$, the transmitter maps a minibatch of $|\mathcal{B}_t|$ randomly chosen uniformly distributed training examples to symbols and then sends them over the channel. The receiver takes the channel observations $\vect{y}_1,\cdots, \vect{y}_{|\mathcal{B}_t|} $ as input and generates $|\mathcal{B}_t|$ probability vectors $f_{\vect{\rho}}(\vect{y}_1), \cdots, f_{\vect{\rho}}(\vect{y}_{|\mathcal{B}_t|})$. Finally, the receiver computes the empirical cross-entropy loss associated with the $|\mathcal{B}_t|$ training examples according to
\begin{align}
    \mathcal{L}_{\mathcal{B}_t} (\vect{\tau}, \vect{\rho})=-\frac{1}{|\mathcal{B}_t|} \sum_{k=1}^{|\mathcal{B}_t|} \log [f_{\vect{\rho}}(\vect{y}_k)]_{m_k},
\end{align}
and the transmitter and receiver parameters are optimized following ~\eqref{eq:SGD}. This training process is repeated iteratively until a certain criterion is satisfied (e.g., a fixed number of training iterations, or a fixed number of iterations during which the loss has not significantly decreased).

\begin{figure*}[t]
    \centering
    \includegraphics[width=1\textwidth]{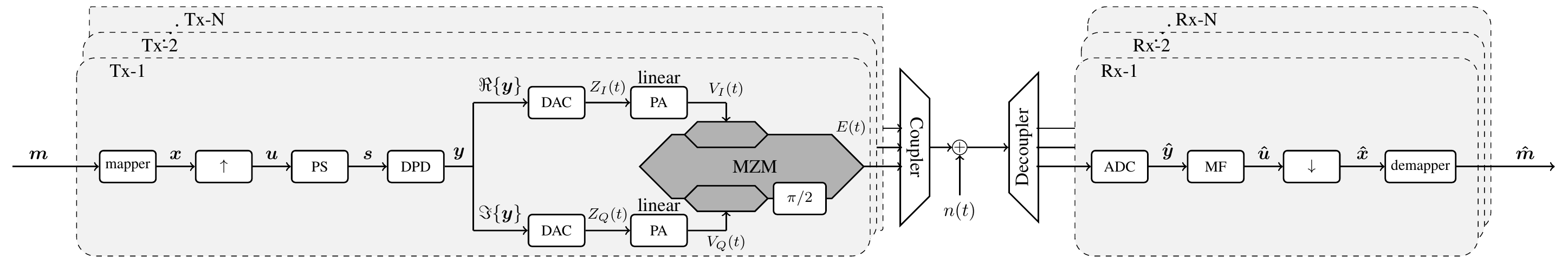}
    \caption{Block diagram showing the conventional \ac{WDM} system ($\uparrow$: upsampling, $\downarrow$: downsampling). The mapper, DPD, and demapper operate on each individual entries of the input sequence, while the \ac{PS} filter operates on a sequence of $2L_1 +1$ signals, where $2L_1 +1$ is the \ac{PS} filter taps. Note, to allow close channel spacing close to the symbol rate, we assume that there are no optical filters or multiplexers.}
    \label{fig:wdm_conventional}
\end{figure*}

\subsubsection{Training Without a Channel Model}
\label{training_without_channel}
In case the channel is unknown or not differentiable, e.g., an experimental channel, the transmitter optimization becomes challenging due to the fact that the gradient of the instantaneous channel transfer function is unknown, thus hindering the numerical computation of the transmitter gradients. One way to circumvent this limitation is to first learn a surrogate channel model, e.g., through supervised learning~\cite{li2020end,wang2020data} or an adversarial process~\cite{ye2018channel, karanov2020concept}, and use the surrogate model to train the transmitter. However, the performance of the resulting system severely degrades if the surrogate model deviates from the real channel. A different approach based on a stochastic transmitter was proposed in~\cite{aoudia2018end, aoudia2019model}. For this approach, the transmitter is regarded as an \ac{RL} agent, and the transmitter and receiver are optimized in an alternating fashion which we review next.

\emph{Receiver training:} 
The receiver training is similar as before. However, this time, the transmitter parameters $\vect{\tau}$ are assumed to be fixed. At each training iteration, the transmitter maps a minibatch of $|\mathcal{B}_t|$ uniformly distributed training examples to symbols and sends them over the channel. The receiver takes the channel observations $\vect{y}_1,\ldots, \vect{y}_{|\mathcal{B}_t|}$ as input and generates $|\mathcal{B}_t|$ probability vectors $f_{\vect{\rho}}(\vect{y}_1),\ldots, f_\rho(\vect{y}_{|\mathcal{B}_t|})$. Then, the receiver takes one optimization step according to $\vect{\rho}_{t+1} = \vect{\rho}_{t} - \alpha \nabla_{\vect{\rho}} \mathcal{L}_{\mathcal{B}_t} (\vect{\tau}_t, \vect{\rho}_{t})$, where $\vect{\tau}_t$ is fixed during receiver training. This training process is repeated iteratively until a certain stop criterion is satisfied.

\emph{Transmitter training:}
For the transmitter optimization, the receiver parameters are assumed to be fixed. At each training iteration, the transmitter performs the symbol mapping as before. In order to allow for the transmitter gradients computation, a small Gaussian perturbation is applied such that $\tilde{\vect{x}} = \vect{x} + \vect{w}$, $\vect{w}\in \mathcal{CN}(0, \sigma^2_p\vect{I}_{N})$, is sent over the channel. Therefore, the transmitter can be interpreted as stochastic and is described by
\begin{align}
    \label{eq:gaussian_policy}
    \pi_{\vect{\tau}}(\tilde{\vect{x}}_k|m_k) = \frac{1}{(\pi \sigma_p^2)^{N}}
    \exp \left(
        - \frac{||\tilde{\vect{x}}_k - f_{\vect{\tau}}(m_k) ||_2 ^2}{\sigma_p^2}
    \right).
\end{align}
Based on the received channel observations, the receiver computes per-example losses $\ell_k=-\log([f_{\vect{\rho}}(\vect{y}_{k})]_{m_{k}})$, and sends them back to the transmitter. Finally, the transmitter parameters $\vect{\tau}$ are updated according to $\vect{\tau}_{t+1} = \vect{\tau}_{t} - \alpha \nabla_{\vect{\tau}} \mathcal{L}_{\mathcal{B}_t}(\vect{\tau}_t)$, where $\nabla_{\vect{\tau}} \mathcal{L}_{\mathcal{B}_t}(\vect{\tau})$ is approximated by
\begin{align}
    \nabla_{\vect{\tau}} \mathcal{L}_{\mathcal{B}_t}(\vect{\tau}) = \frac{1}{N_T}\sum_{k=1}^{N_T} \ell_k \nabla_{\vect{\tau}} \log \pi_{\vect{\tau}}(\tilde{\vect{x}}_k|m_k),  \label{eq:PolicyGradient1}      
\end{align}
for which a theoretical justification can be found in \cite{aoudia2019model}.
Similar to the receiver training, the transmitter learning process is repeated iteratively until a certain stopping criterion is satisfied. Then, the alternating optimization continues again with the receiver learning.

\section{\ac{WDM} System and Main Hardware Limitations}
\label{superchannel}

\subsection{System Model}
Fig.~\ref{fig:wdm_conventional} illustrates the considered \ac{WDM} system. For each channel, a sequence of $|\mathcal{B}_t|$ messages $\vect{m}\in \mathcal{M}^{|\mathcal{B}_t|}$, 
where $\mathcal{M}=\{1,\ldots,M\}$, are mapped individually to constellation points according to a constellation $\mathcal{C}\in \mathbb{C}^M$, to form the sequence of baseband symbols $\vect{x}\in \mathbb{C}^{|\mathcal{B}_t|}$.
The baseband symbols $\vect{x}$ are then upsampled to get $\vect{u}\in \mathbb{C}^{|\mathcal{B}_t|R}$, after which a \ac{PS} filter is applied to get the discrete-time baseband signals $\vect{s}\in \mathbb{C}^{|\mathcal{B}_t| R}$, where $R$ is the upsampling rate.\footnote{An upsampler with $N\times$ upsampling rate increases the sample rate by inserting $N-1$ zeros between samples. Moreover, the transients from the convolution operation, eg., \ac{PS} filtering and matched filtering, are assumed to be removed.} To mitigate the performance degradation caused by the hardware imperfections (in this paper  \ac{ENOB} of the \ac{DAC} and IQM nonlinearity), a \ac{DPD} algorithm is applied. Then, the real and imaginary  part of the pre-distorted signals $\vect{y}\in\mathbb{C}^{|\mathcal{B}_t| R}$ are separately fed to the \acp{DAC} of the in-phase and quadrature branches. Finally, the \acp{DAC} outputs $Z_{I}(t)$ and $Z_{Q}(t)$ are separately amplified to drive the \ac{IQM}, where the driving voltages of the in-phase and quadrature branches are denoted by $V_I (t)$ and $V_{Q}(t)$, respectively. Similar to~\cite{curri2012optimization, khanna2015robust,berenguer2015nonlinear}, the channel model we consider in this paper is restricted to a back-to-back setup, and only \ac{AWGN} with constant power is added to simulate the noise introduced by the booster amplifier.  At the receiver, the received signals are passed through an \acf{ADC}, after which the digitized channel observations $\hat{\vect{y}}\in\mathbb{C}^{|\mathcal{B}_t|R}$ are convolved with a \ac{MF} and then down-converted with rate $R$. Finally, the downsampled signals $\vect{\hat{x}}\in \mathbb{C}^{|\mathcal{B}_t|}$ are individually mapped to the estimates $\hat{\vect{m}}\in \mathcal{M}^{|\mathcal{B}_t|}$ of the transmitted messages.\JS{Note that, as optical filters and multiplexers would prevent close channel spacing due to their finite response, we assume that channels are combined using broadband passive couplers. Thus, there are no optical filters in our system;  such a system is often referred to as  \emph{superchannel} system.}

\subsubsection*{\ac{IQM} Model}
\label{IQM_model}

The coherent optical transmitter used for high-order modulation schemes such as M--QAM, M--PAM is often based on a dual parallel \ac{MZM}.  For an ideal dual parallel MZM biased at the null point, it has been shown that its transfer function becomes~\cite{napoli2018digital}
\begin{align}
    E(t) = E_0 \left[ \sin \left(\frac{\pi V_I (t)}{2 V_{\pi}} \right) + j\sin \left(\frac{\pi V_Q(t)}{2V_{\pi}} \right)\right],
\end{align}
where $E_0$ is the amplitude of the magnitude of the electric field, $V_\pi$ is the required voltage difference to switch ON/OFF the modulator, and $V_I (t)$ and $V_Q (t)$ are the driving voltage of the in-phase and quadrature branches, respectively. 
The intrinsic sinusoidal form of the \ac{MZM} leads to strong signal distortions when driving with a high peak voltage $V_\text{p}$, which must be compensated, e.g., by pre-distortion with an \emph{arcsin} function. Alternatively, one can use a low-driving voltage to operate in the near-linear regime of the modulator. \JS{However, this significantly increases the modulator loss, which results in a degraded optical \ac{SNR} after adding the booster amplifier noise. }

\begin{figure*}[t]
    \centering
    \includegraphics[width=1\textwidth]{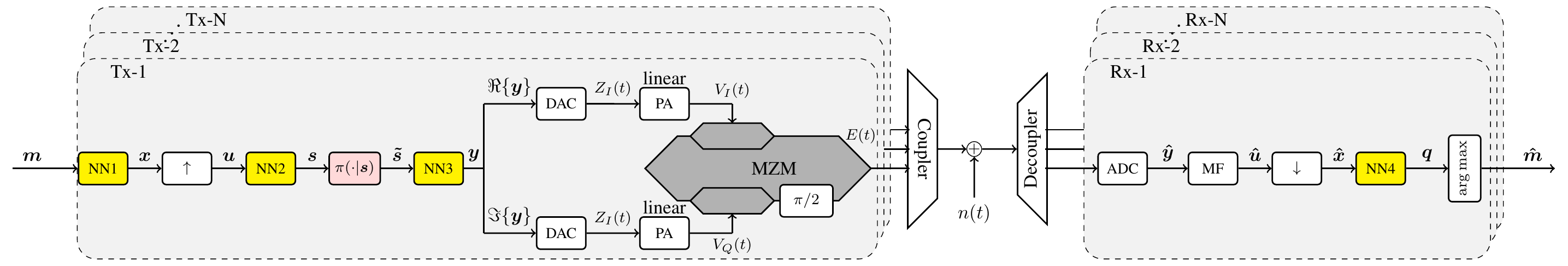}
    \caption{{Block diagram showing the end-to-end AE-learning based \ac{WDM} system ($\uparrow$: upsampling, $\downarrow$: downsampling). The trainable components are highlighted in yellow. NN1, NN3, and NN4 operate separately on each entry of the input sequence, while NN2 takes a vector of length $2L_1 +1$ signals as input, where $2L_1 +1$ is the \ac{PS} filter length. }}
    \label{fig:wdm_AE}
\end{figure*}

\subsubsection*{PA Model}
The \ac{PA} used for amplifying the \ac{DAC} outputs behaves as a nonlinear memory system, i.e., the \ac{PA} output at any time instant depends on the current instantaneous input as well as the inputs at previous time instances. Denoting the memory depth by $L$, the \ac{PA} denoted by $f_\text{PA}: \mathbb{R}^{L+1}\to \mathbb{R}$,  can be defined by
\begin{align}
    V(t) = f_\text{PA}(Z(t), \ldots, Z(t-L)),
\end{align}
where $f_\text{PA}$ is a nonlinear function and $Z(t)$ is the \ac{DAC} output of the real/imaginary branch.
For an ideal \ac{PA} without memory effect, its transfer function becomes $V(t)=GZ(t)$, where $G$ is the \ac{PA} gain.

\subsubsection*{\ac{DAC} Model}
\acp{DAC} used for high-bandwidth optical communications typically have low resolution. Currently, devices on the market provide 8 nominal bits. However, due to the sampling and jitter effects, the noise introduced by quantization is usually enhanced.  One parameter to assess the amount of noise introduced by the \ac{DAC} is the \ac{ENOB}, which is defined as~\cite{laperle2014advances}
\begin{align}
    \text{ENOB} = \frac{ \text{SNDR}(\mathrm{dB}) -1.76}{6.02},
\end{align}
where the \ac{SNDR} is a measurable quantity, and is typically around $35\,$dB. Typically, high-speed \acp{DAC} with 8-bit nominal resolution can be translated into $\text{ENOB}\leq 6$ for operation within the device bandwidth. However, it should be noted that \ac{ENOB} is a varying quantity and it changes over frequency. In this paper, for the sake of simplicity, the \ac{ENOB} is assumed to be constant over the considered bandwidth and is set to 6.\footnote{This is a reasonable assumption for current generation transceivers.} We model the \ac{ENOB} noise introduced by the \ac{DAC} as \ac{AWGN} with variance determined by the \ac{ENOB} of the device~\cite{napoli2016digital}
\begin{align}
    \sigma^2_q = \frac{1}{12}\left(\frac{E_{\text{peak}}}{2^{\text{ENOB}-1}-1}\right)^2,
\end{align}
where $E_\text{peak}=\max( \max(\vert\Re\{\vect{y}\}\rvert), \max(\lvert \Im\{\vect{y}\}\rvert ))$ is the peak amplitude of the input signals. Note that the finite bit-resolution of the \ac{DAC} limits the strength of the \emph{arcsin}-based pre-distortion that can be applied, because it increases the peak amplitude, thereby resulting in higher noise. Therefore, there exists an optimum \ac{DAC} driving voltage which balances \ac{SNR} degradation from \ac{MZM} losses when driving in the linear regime of the modulator and \ac{SNR} degradation from limited compensation of \ac{MZM} nonlinearity when driving at high voltages.

\section{Proposed End-to-End \ac{WDM} System}
\label{proposed_AE}
In this section, we start by introducing the proposed \ac{AE} implementation for the \ac{WDM} system. The symbol rate and modulation formats are assumed to be the same for all channels, and we consider using the same \ac{AE} configurations for all channels.

\subsection{Autoencoder Design}
In principle, the entire transmitter and receiver can be implemented as an \ac{AE} and trained by end-to-end learning as proposed in~\cite{o2017introduction}. However, this leads to:
\begin{enumerate}
        \item[(a)] Difficulty in interpretation: In contrast to conventional communication systems, where the performance of each transmitter/receiver blocks can be measured separately, the \ac{AE} implementation is a ``black-box'', and it is hard to interpret the learned solution and to quantify the origin of the performance improvement.
    \item[(b)] \JS{High training complexity: The transmitter needs to perform several tasks, such as symbol mapping, \ac{PS}, and pre-distortion jointly, and learning the transmitted waveform involves sequential input data, which significantly increases the \ac{NN} size with the ``one-hot'' encoding being applied, therefore increasing the training complexity.}
    \item[(c)] Parameter initialization: It is difficult to know which parameter choice leads to good performance prior to training, and random parameters initialization can slow down or
    even completely stall the convergence process~\cite{mishkin2015all}.
\end{enumerate}
To address these issues, we design our \ac{AE} following the architecture of conventional communication systems as shown in Fig.~\ref{fig:wdm_AE}. The policy $\pi(\cdot|\vect{s})$ can be ignored for now. The transmitter \ac{NN} is decomposed into a concatenation of three simpler (small) \acp{NN}, each corresponding to one functional block of a conventional communication system. By doing this, the parameters of these \acp{NN} can be initialized such that they initially perform close to their conventional counterparts. Moreover, the ``block-wise'' transceiver \ac{NN} design allows for an ablation study and therefore makes it possible to partially explain the learned solution. As a result, the proposed scheme has \emph{decreased training complexity} and \emph{increased interpretability} as compared to a conventional \ac{AE}.\footnote{We note that such an \ac{AE} implementation can potentially lead to performance degradation compared to the conventional \ac{AE}, which we do not study in this paper.}

\begin{table*}[t]
\setlength{\tabcolsep}{0.6em}
\scriptsize
\centering
\vspace{0.15cm}
\caption{NN parameters }
\begin{tabular}{c|c|ccc|ccc|ccc|ccc }
\toprule
 & & \multicolumn{3}{c}{NN1 $f_{\vect{\theta}_1}$} &\multicolumn{3}{|c}{NN2 $f_{\vect{\theta}_2}$}  &\multicolumn{3}{|c}{NN3 $f_{\vect{\theta}_3}$} &\multicolumn{3}{|c}{NN4 $f_{\vect{\theta}_4}$}   \\
 \midrule
 & layer    & input  & hidden& output &input  & hidden & output  &input  & hidden & output &input  & hidden & output      \\ 
\midrule
\multirow{3}{*}{(i)} &{\#~of layers}    &  -  & $3$  & -       & -    & $0$ & -     & -  & $3$ & -  & -  & $2$ & -   \\ 
                     &\#~of neurons   & $M$ & 50   & $2$     & 201  & -   & 1    & -  & $50$ & -  & 2 & $20$ & M  \\ 
                    & act.~function   &  -  & ReLU & Linear  & -    & -    & Linear     & -  & ReLu & -  & -  & ReLu & Softmax   \\ 
\bottomrule
\end{tabular}
\label{tab:network_parameters}
\end{table*}

\subsubsection{Transmitter}
At the transmitter, the symbol mapper, the \ac{PS} filter, and the \ac{DPD} of the conventional communication system are replaced by three \acp{NN}. We denote these three \acp{NN} by $f_{\vect{\theta}_1}(\cdot)$, $f_{\vect{\theta}_2}(\cdot)$, and $f_{\vect{\theta}_3}(\cdot)$, where $\vect{\theta}_1, \vect{\theta}_2$, and $\vect{\theta}_3$ are the sets of trainable parameters. We define these three \acp{NN} in the following:
\begin{enumerate}
    \item[(i)] NN1 $f_{\vect{\theta}_1}$: $\mathcal{M}\to \mathbb{C}$ maps each message $m_k\in \vect{m}$ to a constellation point according to $x_k = f_{\vect{\theta}_1}(m_k)$, where an average power constrain $\mathbb{E}\{|x_k|^2\}=1$ is enforced.
    
     \item[(ii)] NN2 $f_{\vect{\theta}_2}$: $\mathbb{C}^{2L_1+1}\to \mathbb{C}$ generates each of the pulse-shaped baseband signals according to $s_k=f_{\vect{\theta}_2} (\vect{u}_{k}^{(L_1)})$,
     where $\vect{u}_{k}^{(L_{1})}=[u_{k-L_{1}},\ldots, u_{k+ L_{1}}]^{\top}$.
     Here, NN2 only has a single layer applying a linear activation function and can be interpreted as a standard \ac{FIR} filter. Therefore, the generation of the pulse-shaped signal can be described by $s_k=\vect{\theta}_2^\top\vect{u}_k^{(L_1)}$.

     \item[(iii)] NN3 $f_{\vect{\theta}_3}$: $\mathbb{C} \to \mathbb{C}$ generates each of the pre-distorted signals according to $y_{k}=f_{\vect{\theta}_3}(s'_{k})$, 
    where $f_{\vect{\theta}_3}(\cdot)$ operates separately on the in-phase and quadrature branches,
    and $-1\leq \Re{\{s'_k\}, \Im{\{s'_k\}}}\leq1$
    is obtained by normalizing $s_k$ according to $s'_k = s_k/\max\{\max\{\lvert\Re\{\vect{s}\}\rvert\}, \max\lvert\Im\{\vect{s}\}\rvert\} \}$, where $\vect{s}$ is the pulse-shaped signal sequence.
\end{enumerate} 

\subsubsection{Receiver}
At the receiver, only the symbol demapper is replaced by an \ac{NN}, denoted by NN4 $f_{\vect{\theta}_4}$: $\mathbb{C} \to \mathcal{M}$, which maps each of the downsampled signal $y_k$ to the estimate of the transmitted message as described in Section~\ref{AE_Rx}. \JS{We note that, in principle, the \ac{MF} can also be implemented by an \ac{NN}. In a real system, however, the \ac{MF} is usually implemented as part of the adaptive equalizer, and we therefore have left it out of this discussion.}

\subsection{Learning With a Channel Model}
Assuming that all transfer functions of the components in the considered system are known and differentiable, the system can be optimized via standard end-to-end \ac{AE}-learning~\cite{o2017introduction} by minimizing the Monte-Carlo approximation of the cross-entropy loss, defined by
\begin{align}
    \mathcal{L}_{\mathcal{B}_t}(\vect{\theta}_1,\vect{\theta}_2, \vect{\theta}_3,\vect{\theta}_4)=\frac{1}{|\mathcal{B}_t|} \sum_{k=1}^{|\mathcal{B}_t|} \log [f_{\vect{\theta}_4}(y_k)]_{m_k}.
\end{align}
Similar to~\eqref{eq:ce_loss}, the dependence of $\mathcal{L}_{\mathcal{B}_t}(\vect{\theta}_1,\vect{\theta}_2, \vect{\theta}_3,\vect{\theta}_4)$  on $\vect{\theta}_1,\vect{\theta}_2, \vect{\theta}_3$ is implicit through the distribution of the downsampled signal $y_k$, which is a function of the channel input
$g(\tilde{s}_k)$,
where $g(\cdot)$ denotes the joint transfer function of the DAC, PA, and IQM, and $\tilde{s}_k$ is dependent on \acp{NN} 1--3 as can be seen in Fig.~\ref{fig:wdm_AE}. For the optimization, in order to have a faster and more stable convergence, all \acp{NN} are first initialized to mimic their model-based counterparts via pre-training. Then, the sets of parameters $\vect{\theta}_1,\vect{\theta}_2, \vect{\theta}_3,\vect{\theta}_4$ are jointly optimized using the Adam optimizer~\cite{kingma2014adam}.

\begin{algorithm}[t]
\caption{Optimization of the pulse shaping filter}\label{alg:cap}
\begin{algorithmic}[1]
\Repeat\\
\hspace*{\algorithmicindent}\text{$\triangleright$ Transmitter}
    \State Symbol mapping and upsampling: $\vect{m}\to \vect{x} \to \vect{u}$
    \State Pulse shaping: $\vect{u}\to \vect{s}$
    \State Apply Gaussian: $\vect{s} \to \vect{\tilde{s}}$
    \State Apply DPD: $\vect{\tilde{s}}\to \vect{y}$
    \State Send $\vect{y}$ \\
\hspace*{\algorithmicindent}\text{$\triangleright$ Receiver}
    \State Receive: $\vect{\hat{y}}$
    \State Matched filtering and downsampling: $\vect{\hat{y}} \to \vect{\hat{u}} \to \vect{\hat{x}}$
    \State Compute per example loss: $\ell_k$
    \State Send $\ell_k$\\
\hspace*{\algorithmicindent}\text{$\triangleright$ Transmitter}
    \State Receive $\ell_k$
    \State Update NN2 parameters according to \eqref{ps_optimization}
\Until Stop criterion is satisfied
\end{algorithmic}
\end{algorithm}

\subsection{Learning Without A Channel Model}
\label{rl_training}
In practice, training of the proposed \ac{AE} in an experiment is challenging due to the fact that the instantaneous gradients of the physical channel are unknown. To solve this problem, we follow the alternative optimization approach that we reviewed in Section~\ref{training_without_channel}. \JS{The training of the demapper does not require differentiation of the channel can therefore be performed  via supervised learning. For the transmitter training, since the transmitter consists of three \acp{NN}, one can perform the transmitter training by alternating between the optimization of the symbol mapper, the \ac{PS} filter, and the \ac{DPD}.} In this paper however, we only focus on training of the \ac{PS} filter, for which memory effects need to be taken into account. For the optimization of the mapper (i.e., NN1) or the DPD  (i.e., NN3), we refer the reader to~\cite{aoudia2018end, aoudia2019model} and our recent paper~\cite{song2021over}. \JS{To that end, the parameters of the mapper, the DPD, and the demapper (i.e., NN4) are assumed to be pretrained and fixed during the \ac{PS} filter training.}

For the PS filter optimization, the training algorithm described in Section~\ref{training_without_channel} cannot be used directly due to the memory introduced by the matched filtering. Therefore, we extend the training approach as follows. In each training iteration $t$, the transmitter generates a batch of $|\mathcal{B}_t|$ random uniformly distributed messages within one message vector $\vect{m}\in\mathcal{M}^{|\mathcal{B}_t|}$ and maps them individually to the baseband symbols after which $R$-time upsampling is applied. Then, the baseband transmitted signals are generated by convolving the upsampled signals with a real-valued trainable filter according to $\vect{s} = \vect{u}^\top * \vect{\theta}_2^\top$, where $*$ denotes the convolution operator. 
To allow for the gradient computation of the trainable \ac{PS} filter, we consider a Gaussian policy. To that end, a small perturbation $w_k \in \mathcal{CN}(0, \sigma^2)$ is applied to each of the pulse-shaped signals before applying the \ac{DPD}. Therefore, the \ac{DPD} input $\vect{\tilde{s}} = \vect{s}+\vect{w}$ is stochastic and can be described by the PDF
\begin{align}
\label{pdf2}
    \pi_{\vect{\theta}_2}(\tilde{s}_k| \vect{u}_k^{(L2)}) = \frac{1}{\sqrt{2\pi\sigma^2}} e^{-{\frac{|\tilde{s}_k - \vect{\theta}_2^\top \vect{u}_k^{(L_1) }|^2}{2\sigma^2}} }.
\end{align}
At the receiver, the channel observations $\hat{\vect{y}}=[\hat{y}_1,\ldots, \hat{y}_{|\mathcal{B}_t|R}]$ are filtered by a MF and then downsampled with rate $R$. Then, the resulting signals $\vect{\hat{x}}=[\hat{x}_1,\ldots, \hat{x}_{|\mathcal{B}_t|}]$ are used to compute 
the per-example loss defined by 
\begin{align}
    \ell_k = \log[f_{\vect{\theta}_4}(\hat{{x}}_k)]_{m_k}, \quad k=1,\ldots,|\mathcal{B}_t|,
\end{align}
where $\hat{{x}}_k=\hat{u}_{kR}$.
\JS{The per-example losses are sent back to the transmitter to perform the \ac{PS} filter training. Due to memory effects introduced by the convolution operation in the \ac{MF} and \ac{PS} filter, $\ell_k$ is related to a subset of the entire sequence $\vect{x}$ and $\vect{m}$. We denote the total number of samples related to $\ell_k$ by $2G+1$. 
The training objective is to optimize $\vect{\theta}_2$ such that the expected cross-entropy loss $\mathcal{L}(\vect{\theta}_2) = \mathbb{E}\{\ell_k\}$ is minimized.
Following \cite{aoudia2018end, aoudia2019model}, we compute $\nabla_{\vect{\theta}_2}\mathcal{L}(\vect{\theta}_2)$ using the following proposition.

\emph{Proposition 1:} The gradient of $\mathcal{L}_{\mathcal{B}_t}(\vect{\theta}_2)$ can be approximated by 
\begin{align}
    \label{ps_optimization}
    &\nabla_{\vect{\theta}_2} \mathcal{L}_{\mathcal{B}_t}(\vect{\theta}_2)\\ \nonumber
   & \approx\frac{1}{\sigma^{2}}\sum_{g=-G}^{G}\sum_{k=1}^{|\mathcal{B}_{t}|}\frac{1}{|\mathcal{B}_{t}|}\ell_{k}(\hat{\vect{y}},m_{k})(\tilde{s}_{kR+g}-[\vect{f}_{\vect{\theta}_{2}}(\vect{f}_{\vect{\theta}_{1}}(\vect{m}))]_{kR+g})\\
 & \times \nabla_{\vect{\theta}_{2}}[\vect{f}_{\vect{\theta}_{2}}(\vect{f}_{\vect{\theta}_{1}}(\vect{m}))]_{kR+g}, \notag
\end{align}
where we wrote $\vect{f}_{\vect{\theta}_{i}}$ to highlight that the relation is applied to the entire sequence in order to generate  the entire corresponding output. }

\emph{Proof:} See Appendix~\ref{Append}.

\section{Numerical Results}
\label{results}
In this section, we provide extensive numerical results to verify and illustrate the effectiveness of the proposed \ac{AE}-based \ac{WDM} system. The system performance is measured in terms of \ac{SER}, and for all the results presented below, the MF used is the \ac{RRC} filter.

\subsection{Setup and Parameters}
\subsubsection{Simulation setup}
We set $M=64$, and consider a single channel system as well as a \ac{WDM} system with 3 channels. For the 3-channel setup, the guard band between the adjacent channels is $\eta f_{\mathrm{b}}$ (i.e., the channel spacing between neighboring channels is $(1+\eta)f_b$), where  $\eta\geq 0$ and $f_{\mathrm{b}}$ is the symbol rate. The oversampling rate is set to $R=2$ except for part of Section~\ref{3-channel_system}, where we study the impact of the oversampling rate on the performance. Both the \ac{PS} filter and the \ac{MF} have 201 taps. The hardware impairments considered in this paper are restricted to the IQM nonlinearity and the limited ENOB of the DAC, \JS{while the \ac{PA} is assumed to be linear, as the \ac{PA} nonlinearity is negligible when compared to that of the \ac{MZM}.} However, it should be noted that the proposed approach can be readily applied to a more general setup where the other transmitter components are not idealized (e.g., nonlinear PA  and bandwidth-limited \ac{DAC}).   

\subsubsection{Transmitter and Receiver Networks}

Following previous work, all \acp{NN} are implemented as multi-layer fully-connected \acp{NN}, where the ReLU function is chosen as the activation function for the hidden layers. The \ac{NN} parameters used in this paper are summarized in Table~\ref{tab:network_parameters}. 

\subsubsection{Training}
\emph{NN initialization:}
All \acp{AE} are trained by minimizing the end-to-end cross-entropy loss, with the learning rate and batch size set to $0.0002$ and $16000$, respectively.
For the 3-channel setup in particular, we consider using the same \ac{AE} configuration for all 3 channels, and we therefore only minimize the cross-entropy loss of the center channel and then use the parameters of the center channel \ac{AE} for the \acp{AE} of the side channels. All \acp{AE} are trained for $10000$ training iterations. In each training iteration, uniformly distributed training data are randomly generated, and a total number of $1.6\times 10^8$ data samples are used for each \ac{AE} optimization. For the performance evaluation, in order to avoid leakage of training data into the testing set, independent uniformly distributed data are randomly generated for testing.

\subsubsection{Baseline}
\label{baseline}

For the baseline, we use a geometrically shaped constellation that is obtained via training a standard AE~\cite{o2017introduction} over an \ac{AWGN} channel at $\text{SNR}=18\,\mathrm{dB}$.\footnote{\JS{We find that training the standard \ac{AE} at $\text{SNR}=18\,\mathrm{dB}$ leads to a constellation that is more performant than the standard square 64-QAM for a range of SNRs from $0\,\mathrm{dB}$ to $26\,\mathrm{dB}$ over the \ac{AWGN} channel. The SNR of the considered \ac{WDM} system falls into this SNR regime.}}  The PS filter is chosen as the \ac{RRC} filter with roll-off factor $\beta$, which is the same as the MF at the receiver. The DPD, which operates separately on the in-phase and quadrature components, is based on the \emph{arcsin} operation combined with clipping that can be described as~\cite{tang2008coherent}

\begin{align}
  \tilde{s}_k= \begin{cases} 
      \min (\frac{\pi}{2},  V_\text{clip}\, \arcsin(s_k)) & s_k\geq 0 \\
      \max (-\frac{\pi}{2},  V_\text{clip}\,\arcsin(s_k)) & s_k < 0,
   \end{cases}
\end{align}
where the $\arcsin$ linearizes the IQM response, while the  clipping factor $V_\text{clip}$ needs to be optimized  to reduce the peak-to-average power ratio.

\begin{figure}[t]
    \centering
    \includegraphics[width=1\columnwidth]{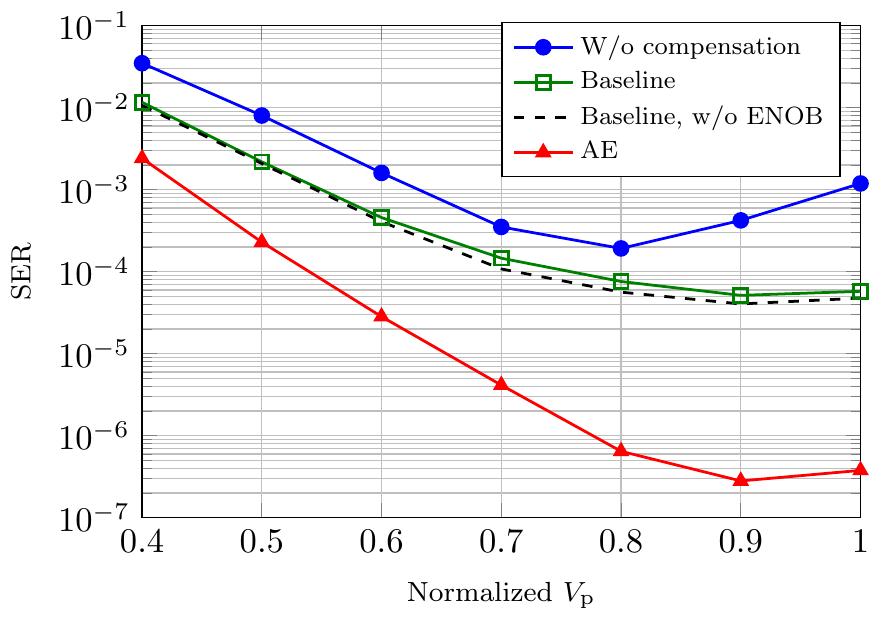}
    \caption{(a):  \ac{SER} performance versus $V_{\text{p}}$ for the single channel scenario with $\beta=10\%$, the blue curve corresponds to the baseline setup but without applying the \emph{arcsin} and clipping based DPD. }
    \label{fig: SER_singel_channel}
\end{figure}

\begin{figure}[t]
    \centering
    \includegraphics[width=1\columnwidth]{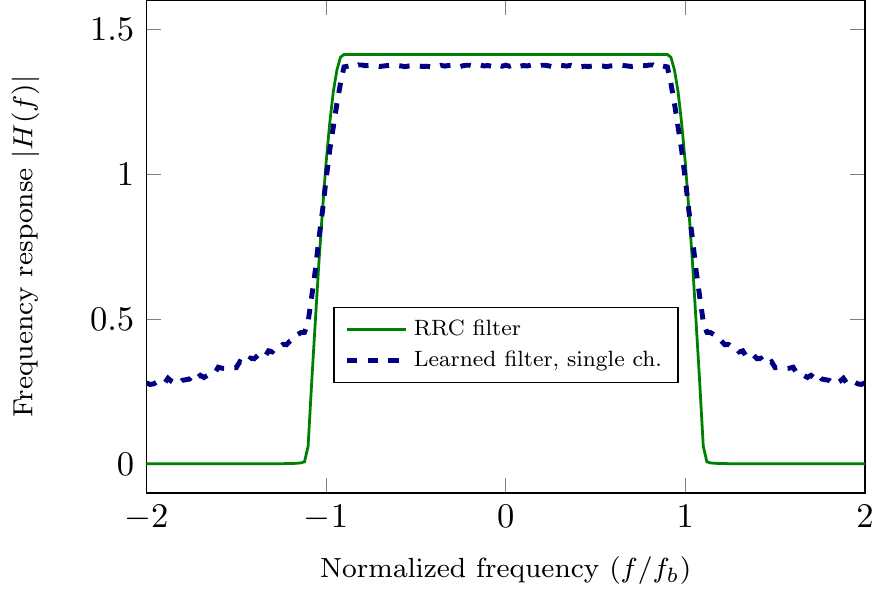}
    \caption{Frequency response of the filter learned in the single-channel setup, showing OOB. The modulator driving swing is $V_\mathrm{p}=1$ and the receiver MF roll-off factor is set to  $\beta=10\%$. The frequency response of the \ac{RRC} filter with $\beta=10\%$ roll-off is also shown as a reference.
    }
    \label{fig:FR_singel_channel}
\end{figure}

\subsection{Results and Discussion}
\begin{figure*}[t]
    \centering
    \includegraphics[width=1\textwidth]{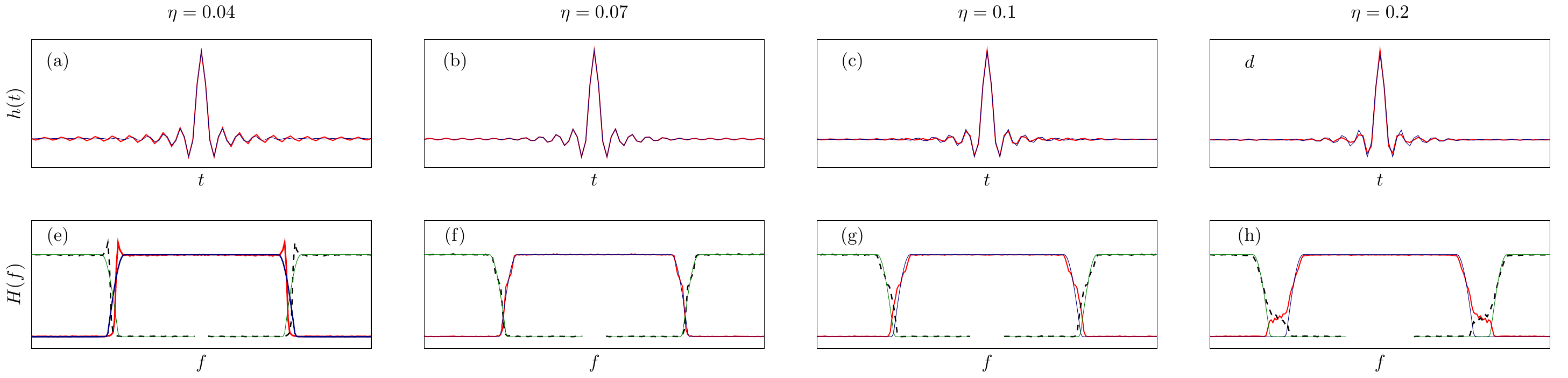}
    \caption{The impulse (top) and frequency (bottom) response of the learned filter (red curve) versus the guard band bandwidth for $R=2$, $V_\text{p}=1$ and $\beta=10\%$. The impulse and frequency response of the RRC filter (blue) with $\beta=10\%$ are also shown as references; The green solid and black dashed curve correspond to the \ac{RRC} filter and the learned filter of the adjacent channels.}
    \label{fig:filter_evolution}
\end{figure*}
\begin{figure*}[t]
    \centering
    \includegraphics[width=1\textwidth]{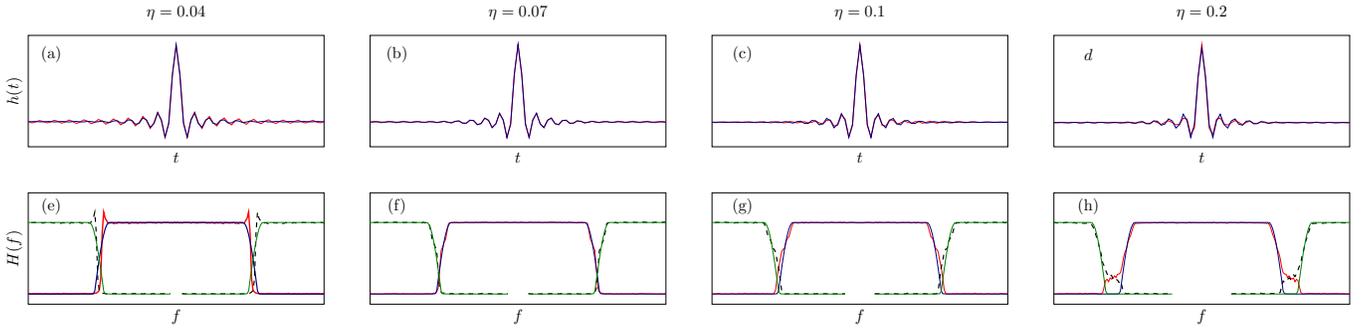}
    \caption{The impulse (top) and frequency (bottom) response of the filters learned in a \ac{WDM} system with 5 channels for $R=4$, $V_\text{p}=1$ and $\beta=10\%$. The impulse and frequency response of the RRC filter (blue) with $\beta=10\%$ are also shown as references.}
    \label{fig:filter_evolution_4x}
\end{figure*}

\subsubsection{Single-Channel System}
We start by investigating a single-channel scenario (e.g., there is no \ac{ICI} in the system), and we evaluate the performance of the proposed method with respect to the peak voltage $V_\text{p}$ of the driving signals. For notation convenience, the peak voltage of driving signals is normalized and the full swing of the \ac{MZM} is used if $V_\text{p}=1$. Due to the dependence of the \ac{MZM} nonlinearity level on the driving voltage swing, a separate \ac{AE} is trained for each considered  $V_\text{p}$.  Fig.~\ref{fig: SER_singel_channel}  visualizes the \ac{SER} of the proposed system when the receiver MF roll-off factor is set to $\beta=10\%$. For a range of considered $V_{\text{p}}$,
the proposed approach achieves significantly better performance than the considered baseline. However, by looking at the frequency response of the learned \ac{PS} filter, as shown with the blue dashed curve in Fig.~\ref{fig:FR_singel_channel}, we observe that compared to the \ac{RRC} filter with $10\%$ roll-off, the learned filter has a significant amount of \ac{OOB} energy, which will introduce severe \ac{ICI} between narrowly-spaced neighboring channels and make it unsuitable for high \ac{SE} \ac{WDM} systems. This result indicates that the system designed for the single-channel setup cannot always be directly applied to a multi-channel setup, and additional care should be taken when designing multi-channel systems.

\subsubsection{\ac{WDM} System With 3 channels}
\label{3-channel_system}
We now train the proposed \ac{AE} in a 3-channel setup.
Fig.~\ref{fig:filter_evolution} visualizes the filters learned with different guard band bandwidth. We start by looking at the impulse response of the learned filters, which appears to be very similar to the \ac{RRC} filter. However, from the frequency responses we observe that the trainable filter learns to adjust its bandwidth according to the guard band between the neighboring channels. In particular, when the guard band is small (e.g., $\eta=0.04$, Fig.~\ref{fig:filter_evolution} (e)) the filter learns to restrict the OOB energy and has a narrower frequency response than the \ac{RRC} filter, indicating that the trainable filter learns to limit \ac{ICI}. As we increase the guard band bandwidth, the bandwidth of the trainable filter increases as well. Similar to the single-channel scenario, the filter learns to put a significant amount of energy in the unoccupied spectrum when the guard band is large (see Fig.~\ref{fig:filter_evolution} (h) for $\eta=0.2$).

\JS{
To train the multi-channel system it is necessary to use high oversampling rates to allow for placing the neighboring channels in the considered spectrum. We emphasize that, in this scenario, it is important to ensure that the \ac{PS} filter cannot generate unrealistically high frequency components. This is illustrated in  Fig.~\ref{fig:filter_evolution_4x}, which depicts the learned filters when the filter is trained with 5 channels and $R=4$ times oversampling rate.
Similar as before, the trainable filter learns to adjust its bandwidth according to the channel spacing. However, the filter also learns to put energy at high frequencies at the edges between the next two channels. Despite this interesting behavior, such a filter is not feasible in practice due to the fact that a practical system would not operate at such high sampling rate because of the hardware limitations as well as power constraints. This result reminds us again the importance of using realistic setups when applying \ac{DL} techniques for designing communication systems. Instead of upsampling to the final oversampling rate before the \ac{PS}, one should use $R=2$ times oversampling rate for the \ac{PS} and another upsampling step after the \ac{PS}, which is the approach we followed for the other multi-channel simulations. An additional benefit of this method is that the number of filter taps is reduced for the same \ac{FIR} filter length, which improves convergence.
}

\begin{figure}[t]
    \centering
    \includegraphics[width=1\columnwidth]{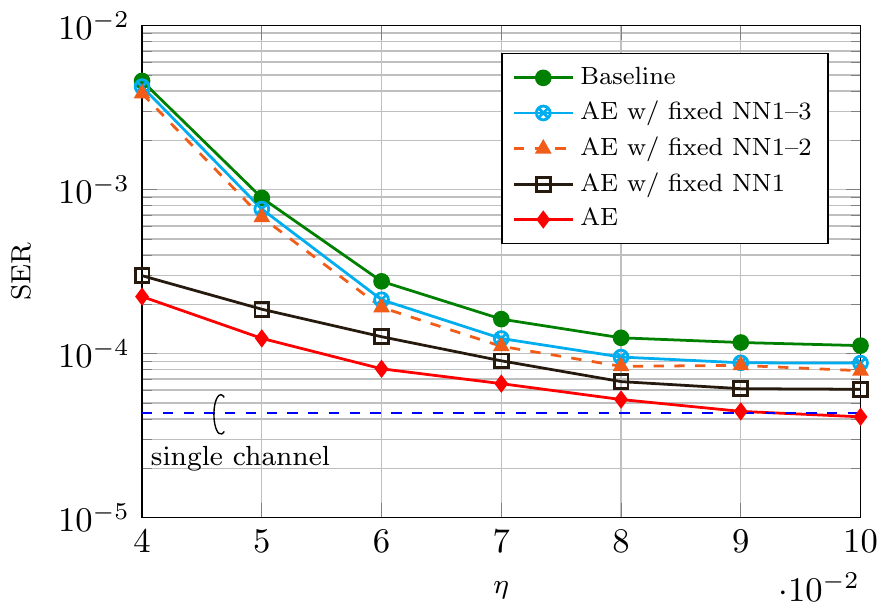}
    \caption{ \ac{SER} performance versus $V_{\text{p}}$ for the  3-channel scenario with $\beta=10\%$, the dashed blue curve corresponds to the baseline scheme for the single-channel scenario.
    }
    \label{fig:3channel-0.1}
\end{figure}

\begin{figure}[t]
    \centering
    \includegraphics[width=1\columnwidth]{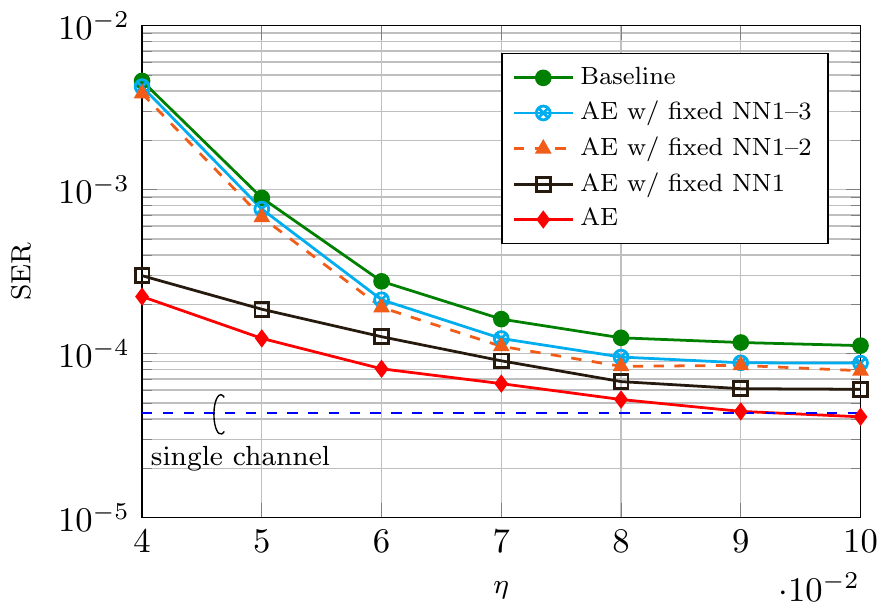}
    \caption{ \ac{SER} performance versus $V_{\text{p}}$ for the  3-channel scenario with $\beta=1\%$, the dashed blue curve corresponds to the baseline scheme for the single-channel scenario.}
    \label{fig:3channel-0.01}
\end{figure}

We now evaluate the performance of the proposed system versus different guard band bandwidth, and we consider setting $R=2$ and the receiver MF roll-off factor to $\beta = 10\%$ and $\beta=1\%$. The achieved  \ac{SER} for the center channel is shown in Fig.~\ref{fig:3channel-0.1} for $\beta=10\%$ and in Fig.~\ref{fig:3channel-0.01} for $\beta=1\%$.\footnote{We note that the side channels have better  \ac{SER} performance than the center channel as they suffer from less ICI.} As a reference, the \ac{SER} performance of the baseline scheme applying \emph{arcsin} combined with clipping is also shown. 
We remark that the clipping factor $V_\text{clip}$ and $V_{\text{p}}$ are optimized for the baseline scheme, while $V_{\text{p}}$ is set to $1$ in the proposed scheme for simplicity. Potentially, the performance of the proposed scheme can be further improved by optimizing $V_{\text{p}}$\textemdash the optimal performance for the single-channel case is achieved at $V_{\text{p}}=0.9$ (see Fig.~\ref{fig: SER_singel_channel}).
For roll-off factors of $10\%$ (Fig.~\ref{fig:3channel-0.1}) and $1\%$ (Fig.~\ref{fig:3channel-0.01}), the proposed  approach outperforms the baseline scheme over all considered guard bands. More importantly, compared to the baseline scheme, the guard band for the proposed scheme can be significantly reduced with limited impact on the \ac{SER} performance
\textemdash for the target \ac{SER} where the baseline performance starts to saturate, the guard band can be reduced by around $37\%$ for $10\%$ roll-off and around $50\%$ for $1\%$ roll-off. Such results indicate that the proposed approach can improve the \ac{SE} of \ac{WDM} systems by allowing to put the channels at a very narrow channel spacing. \JS{However, it should be noted that the reduction in guard bands does not translate directly into the same gain in terms of \ac{SE}, as the explicit \ac{SE} depends on the applied modulation formats, the channel spacing, and the resulting \ac{SER}}.

\begin{figure}[t]
    \centering
    \includegraphics[width=0.76\columnwidth]{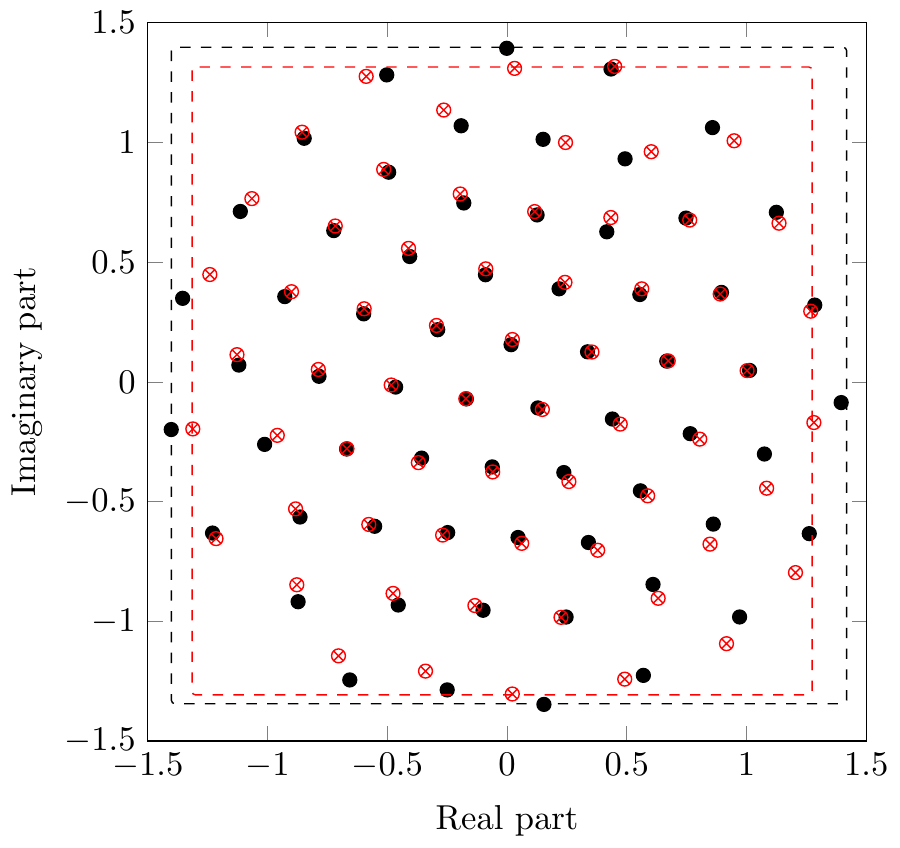}
    \caption{Constellation used for the baseline (black) and constellation learned in the 3-channel setup with $\eta=0.05$ (red). 
    }
    \label{fig:constellations}
\end{figure}

\begin{figure}[t]
    \centering
    \includegraphics[width=1\columnwidth]{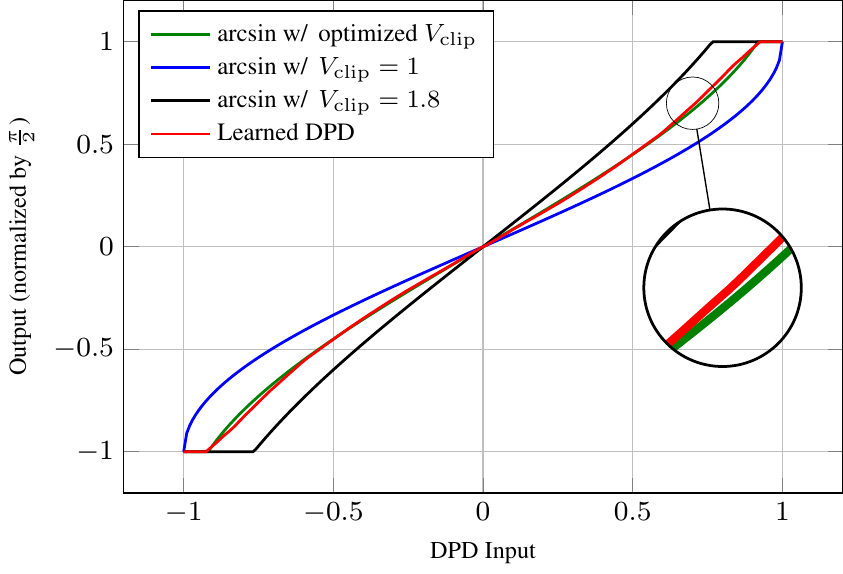}
    \caption{
    Transfer function of NN3 and the conventional DPD using \emph{arcsin} combined with optimized clipping for $\beta=10\%$ and $\eta=0.05$. The transfer functions of the conventional DPD using \emph{arcsin} and sub-optimal clippings are also shown as references. }
    \label{fig:nonlinear_func}
\end{figure}

\subsubsection{Learned Constellation}
\label{learned_const}
Fig.~\ref{fig:constellations} visualizes the learned constellation when the \ac{AE} is trained in the 3-channel setup with $\beta=10\%$ and $\eta=0.05$. The constellation optimized over the \ac{AWGN} channel and used for the baseline is also shown as a reference. It is shown that the constellation optimized over the \ac{WDM} setup has lower peak amplitude than the baseline, indicating the constellation optimized for the \ac{AWGN} channel is suboptimal for a system that is impaired by hardware imperfections. One possible explanation for such observation is that the \ac{AE} learns to limit the peak voltage $V_p$  by restricting the maximum amplitude of the constellation, so as to limit the signal distortion caused by the nonlinear \ac{MZM}.

\subsubsection{Learned DPD}
Fig.~\ref{fig:nonlinear_func} visualizes the transfer function of the \ac{DPD} (i.e., ANN3) learned for the 3-channel system with $\beta=10\%$ and $\eta=0.05$. The transfer functions of the conventional \ac{DPD} employing \emph{arcsin} and different clipping  $V_\text{clip}$ are also shown as references. It is shown that the baseline \ac{DPD} with optimized clipping has a  response similar to the learned \ac{DPD}, suggesting that the considered DPD applying \emph{arcsin} combined with optimized clipping is near optimal for the considered scenario.

\subsubsection{Ablation study}
In order to quantify the origin of the performance gains,
we carry out an ablation study by first freezing all the pre-trained NNs and then individually unfreezing them in the order of NN4, NN3, NN2, and NN1. 
We start by unfreezing NN4. The resulting \ac{SER} performance for $\beta = 10\%$ and $\beta = 1\%$ is shown in Fig.~\ref{fig:3channel-0.1}  and Fig.~\ref{fig:3channel-0.01}, respectively. Compared to the baseline scheme, it can be seen that the proposed approach achieves slightly better performance. Such result is what one would have expected, as the demampper trained over the \ac{AWGN} channel is likely to be suboptimal for a channel impaired by hardware imperfections. We then further unfreeze the parameters of NN3 (i.e., NN1--2 are frozen). The resulting performance is very similar (slightly better) to the case where NN1--3 are frozen. This result is consistent with what is shown in Fig.~\ref{fig:nonlinear_func}.  Finally, the parameters of NN2 are also made trainable (i.e., only NN1 is fixed). In this case, the \ac{SER} of the proposed approach improves significantly. Particularly,  the largest gain achieved for $\beta=10\%$ is $\eta = 0.04$ while is $\eta = 0.01$ for $\beta=1\%$, indicating that the guard band can be optimized to improve the system performance.  Finally, when all \acp{NN} are made trainable, the performance of the proposed method further improves, which is consistent with what is shown in Section.~\ref{learned_const}.

\subsection{Model-Free Training of the Pulse-Shaping Filter }
In this section, we extend our results to the case where a differentiable channel model is unknown. Here, we only consider training of the PS filter with the generalized training algorithm discussed in Section \ref{rl_training}. The reason for only learning the \ac{PS} filter is that \ac{PS} filter training contributes to most of the performance gain as it is shown in the ablation study. \ac{RL}-based training of the mapper and the DPD can be found in ~\cite{aoudia2018end, aoudia2019model}, and~\cite{song2021over}, respectively.

\begin{figure}[t]
    \centering
    \includegraphics[width=1\columnwidth]{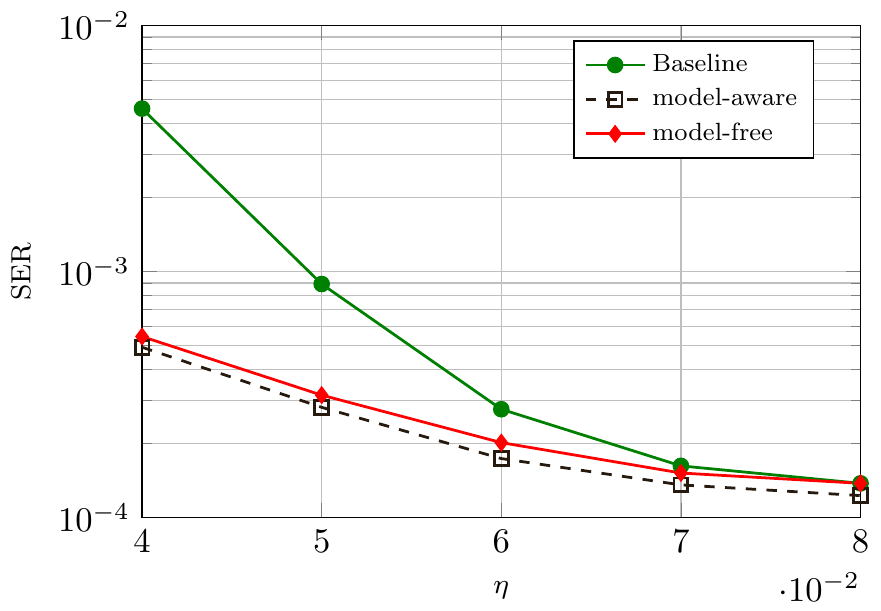}
    \caption{ \ac{SER} performance comparison for $\beta=10\%$ when the proposed AE is trained with and without the perfect channel knowledge.}
    \label{fig:rl_vs_sl}
\end{figure}

Fig.~\ref{fig:rl_vs_sl} shows the  achieved \ac{SER} of the different schemes over a 3-channel \ac{WDM} system. It is observed that the learned \ac{PS} filter using the RL-based algorithm
achieves very similar performance to the one using standard end-to-end learning assuming perfect channel knowledge. However, it should be noted that the \ac{RL}-based approach allows for training of \acp{NN} in an experimental channel, and it has the potential to exceed the performance of the conventional end-to-end learning-based approach, as the performance of the latter is highly dependent on the accuracy of the model used for training.

\section{Conclusion and Future Work}
\label{conclusion}

We proposed a novel end-to-end \ac{AE} for \ac{WDM} systems that are impaired with non-ideal hardware components. In contrast to most of the conventional \acp{AE}, which are usually implemented as a pair of \acp{NN}, our \ac{AE} design follows the architecture of conventional communication systems, and our transmitter is implemented by a concatenation of simple \acp{NN}. Simulation results show that the proposed \ac{AE}-based system achieves significantly better performance than the considered baseline, and allows to increase the \ac{SE} of \ac{WDM} systems by reducing the channel spacing without severe \ac{SER} performance degradation. By means of an  ablation study, we  quantify the origin of the performance improvement. It is shown that the performance gain can be ascribed to the optimized constellation mapper, \ac{PS} filter, and demapper.  In addition, in case the channel model is unknown, we have shown that the \ac{PS} filter can be trained using \ac{RL}, and our simulation results indicate that the extended \ac{RL}-based training approach can achieve similar performance to the standard end-to-end learning assuming perfect channel knowledge.

For future work, there are several important aspects concerning the use of AEs which deserve further study:
\begin{itemize}
    \item Channel models: We have considered an optical back-to-back channel due to the fact that the hardware distortions alone significantly degrade the system. However, practical systems further suffer from performance loss caused by the nonlinear crosstalk between adjacent channels. The \ac{AE}-based method may help to reduce the impact of the crosstalk and provide significant performance improvement.
    
    \item The current \ac{AE} design assumes that the \ac{WDM} channels operate at the same rate. Practical systems, however, allow for transmission at different rates. New \ac{AE} design and training methods may be needed to allow for flexible transmission rates. 
\end{itemize}

\appendix
\label{Append}

We work with
complete sequences, so that the loss is given by:
\begin{align}
\mathcal{L}(\vect{\theta}_2) & =\mathbb{E}_{\vect{m},\vect{s},\tilde{\vect{s}},\vect{y},\hat{\vect{y}},\hat{\vect{x}}}\{\ell_{k}\}\\
 & =\mathbb{E}_{\vect{m},\tilde{\vect{s}}|\vect{m},\hat{\vect{y}}|\tilde{\vect{s}}}\{\ell_{k}\} \nonumber
\end{align}
where in the second step we remove all the deterministic relations.
Hence 
\begin{align}
& \mathcal{L}(\vect{\theta}_2)  =\sum_{\vect{m}}\iint p(\vect{m})p(\tilde{\vect{s}}|\vect{m})p(\hat{\vect{y}}|\tilde{\vect{s}})\ell_{k}(\hat{\vect{y}},m_{k})\mathrm{d}\tilde{\vect{s}}\mathrm{d}\hat{\vect{y}}\\
 & =\sum_{\vect{m}}\iint p(\vect{m})\pi(\tilde{\vect{s}}|\vect{f}_{\vect{\theta}_{2}}(\vect{f}_{\vect{\theta}_{1}}(\vect{m})))p(\hat{\vect{y}}|\tilde{\vect{s}})\ell_{k}(\hat{\vect{y}},m_{k})\mathrm{d}\tilde{\vect{s}}\mathrm{d}\hat{\vect{y}}, \nonumber
\end{align}
where we wrote $\vect{f}_{\vect{\theta}_{i}}$ to expressly denote that the relation is applied to the entire sequence in order to generate entire the corresponding output. 
Exploiting the policy gradient theorem ~\cite{aoudia2019model}
 and using the fact that $\nabla_x\log(g(x))=\frac{\nabla_x g(x)}{g(x)}$, it then follows that 
\begin{align*}
& \nabla_{\vect{\theta}_{2}}\mathcal{L}(\vect{\theta}_2) \\
& =\sum_{\vect{m}}\iint p(\vect{m})\nabla_{\vect{\theta}_{2}}\pi(\tilde{\vect{s}}|\vect{f}_{\vect{\theta}_{2}}(\vect{f}_{\vect{\theta}_{1}}(\vect{m})))p(\hat{\vect{y}}|\tilde{\vect{s}})\ell_{k}(\hat{\vect{y}},m_{k})\mathrm{d}\tilde{\vect{s}}\mathrm{d}\hat{\vect{y}}\\
 & =\mathbb{E}\{\ell_{k}(\hat{\vect{y}},m_{k})\nabla_{\vect{\theta}_{2}}\log\pi(\tilde{\vect{s}}|\vect{f}_{\vect{\theta}_{2}}(\vect{f}_{\vect{\theta}_{1}}(\vect{m})))\}\\
 & =\mathbb{E}\{\ell_{k}(\hat{\vect{y}},m_{k})\sum_{i=1}^{R|\mathcal{B}_{t}|}\nabla_{\vect{\theta}_{2}}\log\pi(\tilde{s}_{i}|[\vect{f}_{\vect{\theta}_{2}}(\vect{f}_{\vect{\theta}_{1}}(\vect{m}))]_{i})\}\\
 & \approx\mathbb{E}\{\ell_{k}(\hat{\vect{y}},m_{k})\sum_{g=-G}^{G}\nabla_{\vect{\theta}_{2}}\log\pi(\tilde{s}_{kR+g}|[\vect{f}_{\vect{\theta}_{2}}(\vect{f}_{\vect{\theta}_{1}}(\vect{m}))]_{kR+g})\}\\
 & =\frac{1}{\sigma^{2}}\sum_{g=-G}^{G}\mathbb{E}\{\ell_{k}(\hat{\vect{y}},m_{k})(\tilde{s}_{kR+g}-[\vect{f}_{\vect{\theta}_{2}}(\vect{f}_{\vect{\theta}_{1}}(\vect{m}))]_{kR+g})\\
 & \times \nabla_{\vect{\theta}_{2}}[\vect{f}_{\vect{\theta}_{2}}(\vect{f}_{\vect{\theta}_{1}}(\vect{m}))]_{kR+g}\}\\
 & \approx\frac{1}{\sigma^{2}}\sum_{g=-G}^{G}\sum_{k=1}^{|\mathcal{B}_{t}|}\frac{1}{|\mathcal{B}_{t}|}\ell_{k}(\hat{\vect{y}},m_{k})(\tilde{s}_{kR+g}-[\vect{f}_{\vect{\theta}_{2}}(\vect{f}_{\vect{\theta}_{1}}(\vect{m}))]_{kR+g})\\
 & \times \nabla_{\vect{\theta}_{2}}[\vect{f}_{\vect{\theta}_{2}}(\vect{f}_{\vect{\theta}_{1}}(\vect{m}))]_{kR+g},
\end{align*}
which leads us to \eqref{ps_optimization}. The first approximation considers that $\ell_{k}(\hat{\vect{y}},m_{k})$
is only affected by $2G+1$ surrounding samples, while the second
approximation is used to compute the expectation by averaging over
the batch. We ignored boundary effect at the start and end of the sequence.

\balance

\bibliographystyle{IEEEtran}
\bibliography{references} 
\end{document}